\documentclass[aps,pre,reprint,notitlepage,twocolumn,longbibliography]{revtex4-2}
\usepackage{graphicx}
\usepackage{amsmath}
\usepackage{color}
\usepackage{soul}
\graphicspath{{./figs/}}

\def\re {R_\lambda}
\def\recrit {\hat{R}_{\lambda,tr}}
\def\retr {R_{\lambda,tr}}

\def\avdis {\la\epsilon\ra}
\def\la {\langle}
\def\ra {\rangle}

\def\be {\begin{equation}}
\def\ee {\end{equation}}

\newcommand{\rfig}[1] {Fig.~\ref{#1}}

\begin{document}

\title{Emergence of universal scaling in isotropic turbulence}
\author{Sualeh Khurshid}
\affiliation{Department of Mechanical Engineering, Massachusetts Institute of Technology}
\affiliation{Department of Aerospace Engineering, Texas A\&M University}
\author{Diego A. Donzis}    \affiliation{Department of Aerospace Engineering, Texas A\&M University}
\author{Katepalli R. Sreenivasan}
\affiliation{Department of Mechanical \& Aerospace Engineering, Department of Physics, Courant Institute of 
Mathematical Sciences, New York University}

\begin{abstract}
Universal properties of turbulence have been associated traditionally with very high Reynolds numbers, but recent work has shown that the onset of the power-laws in derivative statistics occurs at modest microscale Reynolds numbers of the order of 10, with the corresponding exponents being consistent with those for the inertial range structure functions at very high Reynolds numbers. In this paper we use well-resolved direct numerical simulations of homogeneous and isotropic turbulence to establish this result for a range of initial conditions with different forcing mechanisms. We also show that the moments of transverse velocity gradients possess larger scaling exponents than those of the longitudinal moments, confirming past results that the former are more intermittent than the latter.
\end{abstract}

\maketitle

\section{Background}
Turbulence is characterized by strong amplitude fluctuations over spatial scales that range, nominally, from large scales $O(L)$
to dissipating small scales 
$O(\eta)$. A similar range of temporal scales exists as well.
Within the classical turbulence phenomenology,
large scales depend on the geometry of the flow, or the
generation mechanism of turbulent fluctuations,
while the small scales increasingly approach universal behavior as the scale-separation grows \cite{K41,MY.II,batchelor1953,K62,frisch95,SA97}.
A measure of scale-separation in turbulent flows
is the Reynolds number. The universality of
small scales at high Reynolds numbers is an enduring
notion in turbulence theory and forms the
bedrock of most modelling approaches \cite{K41,MY.II,SheLeveque1994,SA97,sreenivasanyakhot2021,yakhot1992}. 
This view was 
formalized in 
Kolmogorov's
seminal work (K41) and its subsequent modifications (see \cite{K62,frisch95,SA97}), which characterize the statistical behavior
of fluctuations at different scales, in particular
a Reynolds-number-independent state of the
inertial range $(L \gg r \gg \eta)$.
Universality
of small scales is expected to manifest more rigorously at high Reynolds numbers,
which has been the motivation for studying turbulence
at ever increasing Reynolds numbers.

The theory itself does not provide guidance on how high a Reynolds number should be regarded as ``high enough", but it is not uncommon to regard an $\re$ of the order of a few hundred or more as necessary for the inertial scaling to appear \cite{SA97,BenziEtAl1993,BenziEtAl1995,IyerEtAl2017,IyerEtAl2020,tsuji2004,IGK2009,DS2010b,Corrsin1958}. Here, 
$$\re\equiv\sqrt{5/(3 \langle \epsilon \rangle \nu)}u_{rms}^2,$$ 
where $ \epsilon = 2\nu  s_{ij}s_{ij}$ 
is the (instantaneous) energy dissipation rate, $\nu$ is the kinematic viscosity and $u_{rms}$ is the root-mean-square fluctuating velocity; $s_{ij}$ is the rate of strain
given by $(\partial_i u_j + \partial_j u_i)/2$
using Einstein's summation convention 
and angular brackets 
indicate volume averages over an ensemble of realizations in time.
However, recent work has shown that moments of the longitudinal
velocity gradient, which characterizes 
small-scale activity, transitions to power-law scaling at much lower $\re$ than expected
from standard scaling arguments, with the same universal exponents that characterize high Reynolds numbers \cite{SchumacherEtAl2014}. For fluctuations
forced by Gaussian white noise at large scales,
it was also shown
that the transition
from the Gaussian state to the scaling state occurs at 
very low Reynolds number, $\re$ = O(10)
\cite{YakhotDonzis2017,YakhotDonzis2018}. 

\section{Motivations and goals}
\begin{figure}
\includegraphics[trim={2cm 10cm 10cm 6cm},clip,width=1.0\linewidth]{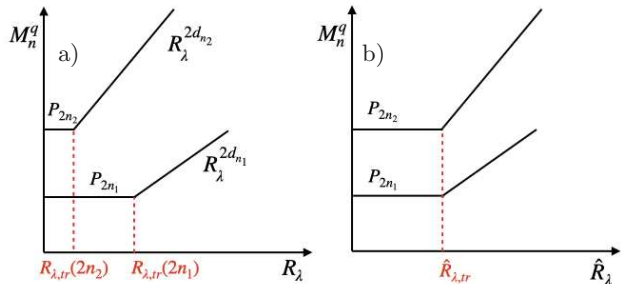}
\begin{picture}(0,0)
\put(-100,95){a)}
\put(15,95){b)}
\end{picture}
\caption{\label{cartoon}
Sketch depicting the transition from the low-$\re$ Gaussian asymptote to power-law scaling
for moments of two different orders $(n_2>n_1)$ 
with respect to (a) the standard Reynolds number, and (b) order-independent Reynolds number.
}
\end{figure}

This result, if true, is important because it implies that the inertial range properties are incipient even when the actual inertial range does not exist in its full splendor. It is unclear from previous studies whether the onset of power-law scaling, as well as its exponents, depend on the particular large-scale forcing used to generate turbulence.
If the results are independent of the forcing, they provide added credibility for the notion of universal scaling at low $\re$. So we drive the system
at large scale using different forcing schemes and test the universality of the transition of
velocity gradients and their scaling as discussed in \cite{YakhotDonzis2017,YakhotDonzis2018}. Specifically, Yakhot \& Donzis
\citep{YakhotDonzis2017} showed that even-order moments of
longitudinal velocity gradients 
\begin{eqnarray}
M_{2n}^{||}=\langle(\partial_{\alpha} u_{\alpha})^{2n}\rangle/\langle(\partial_{\alpha} u_{\alpha})^{2}\rangle^n
\end{eqnarray}
exhibit
Gaussian behavior below a critical $\retr(n)$, beyond
which an order-dependent power-law scaling 
is observed---that is, $M_{2n}^{||}\propto \re^{2d_n}$ for $\re > \retr(n)$. 
The transition Reynolds number $R_{\lambda,tr}(n)$ depends on the moment order as $R_{\lambda,tr}(n)\propto \recrit^{{2nd_n\over 2d_n+3}}$, where the order-independent Reynolds number 
$\recrit(n)\equiv L^2 {\langle(\partial_{\alpha}u_{\alpha})^n\rangle}^{1\over n}/\nu$
is of order 10, though
slowly decreasing with increasing $n$ \cite{YakhotDonzis2018}. 
Higher-order moments transition to power-laws
at lower Reynolds numbers, presumably because they capture stronger fluctuations (which are rare). 
This is sketched in Fig.~\ref{cartoon}.
If the low-$\re$ asymptote (denoted by $P_{n}$) as well as 
the transition Reynolds number are known, the (high-$\re$) scaling exponents 
can be deduced simply by matching the two asymptotes at the transition Reynolds number (Fig.~\ref{cartoon}).
Indeed, the scaling exponents so obtained are
\begin{equation}
\begin{split}
& d_{n}= -{2 n \log (\recrit)-3 n C'+2 \log (P_{2n})\over 4 C'} +\\
& {\sqrt{(2 n \log (\recrit)+3 n C'-2 \log (P_{2n}))^2+24 n C' \log (P_{2n})} \over 4C'}
\end{split}
\label{dneqn}
\end{equation}
where $C'=\log (C)$ and $C$ is a constant of about 90 \cite{YakhotDonzis2017,YakhotDonzis2018}. 
As already stated, these derivative exponents are consistent with the structure function exponents in the inertial range \cite{SchumacherEtAl2014}, obtained from 
experiments and simulations at high Reynolds numbers. Thus, although
there is no inertial range near Reynolds numbers marking the
transition \cite{SchumacherEtAl2007a,YakhotSreenivasan2005,YakhotDonzis2018,sreenivasanyakhot2021},
its signature is apparently present already at very modest $\re$. 

A related question is whether 
longitudinal and transverse gradients scale differently
\cite{ChenEtAl1997,DonzisEtAl2008,GotohEtAl2002,IyerEtAl2020,gotoh2022},
which we assess here within the context of the transition
from Gaussian fluctuations to those with power-law scaling.
We also study gradient-dependent quantities such as 
the enstrophy density $\Omega\equiv\omega_i\omega_i$,
where $\boldsymbol{\omega}=\nabla\times \boldsymbol{u}$ is vorticity, as well as energy dissipation, 
$\epsilon = 2 \nu {s_{i,j} s_{i,j}}$.
Enstrophy is more intermittent than dissipation \cite{DonzisEtAl2008},
but their mean values are related as $\avdis= \nu\la\Omega\ra$.
The scaling of enstrophy is important for understanding
finite-time blow-up and uniqueness problems of fluid-dynamical
equations \cite{GD05, beale1984}.
We relate the new findings for enstrophy to the recent work on the transition in the nature of dissipation
\cite{YeungEtAl2012,SchumacherEtAl2007,SchumacherEtAl2014,PandeySchumacher2020}.
\vspace{-0.5cm}
\section{Direct Numerical Simulations}
We study homogeneous, isotropic
turbulence in a triply periodic domain governed by forcing the incompressible Navier-Stokes equations
\begin{eqnarray}
{\partial u_i\over \partial x_i}&=& 0 
\label{eq:NS1} \\
{\partial u_i\over \partial t} + u_j{\partial u_i\over\partial x_j} &=& -{1\over\rho} {\partial p\over\partial x_i}+\nu{\partial^2 u_i\over\partial x_i^2} + f_i
\label{eq:NS2}
\end{eqnarray}
where $u_i$ is the velocity component in the $x_i$ direction,
and $p$ and $\nu$ are pressure and viscosity, respectively. The forcing term $f_i$ adds energy
into the system to balance dissipation and achieve a statistically stationary
state. 
Details of different forcing mechanisms are 
described next and summarized in table \ref{forcings}.

Gaussian forcing is widely
	used in the literature \cite{EswaranPope1988,YakhotDonzis2017,YakhotDonzis2018,gotoh2022}. We use three different forcing schemes $\boldsymbol{f}$. 
	First, we modify the forcing such that forcing amplitudes 
have an exponential distribution with a random phase, and term this as exponential forcing. Both Gaussian and exponential forcing schemes are white in time.
Second, we
use the so-called linear forcing in which the 
forcing is proportional to the velocity field, that is, 
$\boldsymbol{f}=A\boldsymbol{u}$ where $A$ is a constant.
This forcing has qualitative resemblance to that experienced
in a turbulent flow subjected to mean shear
\cite{LinkmannMorozov2015,McCombEtAl2015,ShihEtAl2015,BarkleyEtAl,SchumacherEtAl2007}. 
As with the stochastic forcing this forcing is applied at low wavenumbers \cite{IGK2009,McCombEtAl2015,ShihEtAl2015}. 
Third, we implemented a modification
by forcing the momentum equations
with vorticity (i.e.\ $\boldsymbol{f}=A\boldsymbol{\omega}$),
which is intermittent unlike the velocity, but do not present these results here. 
We note that all the forcing functions are limited to low wavenumbers and exhibit nearly Gaussian statistics in physical space. However, their dynamics differ qualitatively.

Equations (\ref{eq:NS1}) and (\ref{eq:NS2}) are solved using 
a standard psuedospectral method \cite{rogallo,donzisphd}
with very good small scale resolution $k_{max}\eta\gtrsim 3$. The time step
is evolved using a second-order Runge-Kutta algorithm with a constant
time step such that the Courant-Friedrichs-Lewy condition
$(CFL=|u_{max}|\Delta t/\Delta x)$ remains
below 0.3. These high resolutions allow us to reliably
measure higher order moments of velocity gradients
\cite{YeungEtAl2018,BuariaEtAl2019}. 
All simulations are initialized with 
the same velocity field. To guarantee convergence, we record {\it at least} 50 large scale
eddy turnover times in the stationary state. Gradient moments are computed using
at least 100 snapshots separated by about half an eddy turnover time. 
We have
verified that the skewness in the scaling range is -0.5, the ratio of longitudinal and transverse integral length scales is 2, and that the kinematic constraint 
between the longitudinal correlation function $f(r)=\langle u_{\alpha}(x_{\alpha})u_{\alpha}(x_{\alpha}+r)\rangle/\langle u_{\alpha}^2\rangle$ and the transverse correlation function $g(r)=\langle u_{\beta}(x_{\alpha})u_{\beta}(x_{\alpha}+r) \rangle/\langle u_{\beta}^2\rangle$, 
$\beta$ being orthogonal to $\alpha$, namely $g(r)=f(r)+(r/2)f'(r)$, is satisfied accurately.

\begin{table}
\centering
\caption{\label{forcings}Details of forcing. Gauss and Exponential are stochastic forcing schemes that are white in time and follow
Gaussian and exponential distributions, respectively. 
$A$ is a constant. 
The maximum and minimum 
small-scale resolution in units of $k\eta$ are 60 and 3, depending on $R_\lambda$, for all types of forcing.
}

\begin{ruledtabular}
\begin{tabular}{rlll}
Type & Forcing band \\
Gauss & $0< k\leq 2$ \\
Exponential & $0< k\leq 2$ \\
$u(k)$ & $5\leq k\leq 6$ \\
$\omega(k)$& $5\leq k\leq 6$\\

\end{tabular}
\end{ruledtabular}

\end{table}

\section{Asymptotic states and scaling}
\begin{figure}
\centering
\includegraphics[width=.9\linewidth]{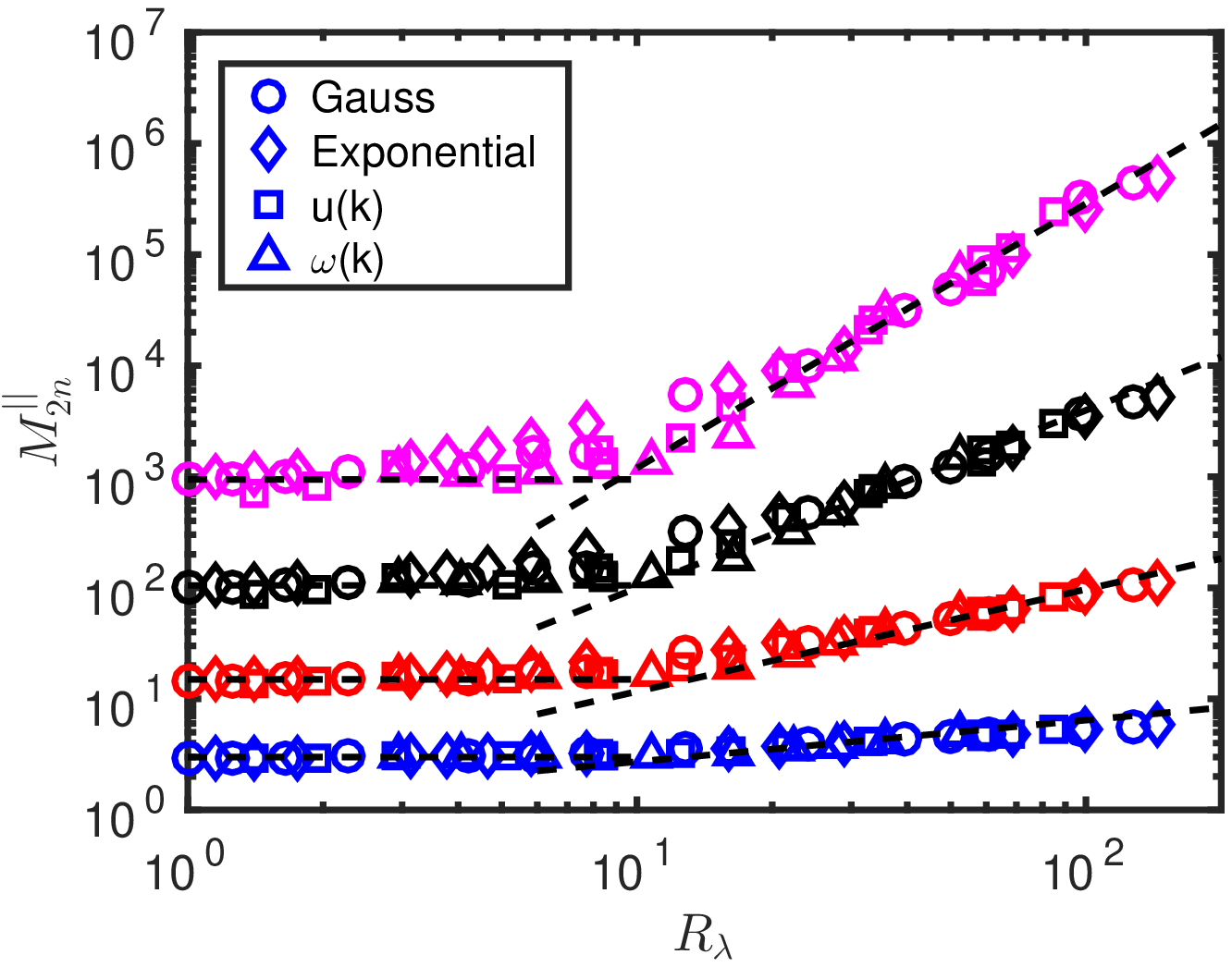}
\includegraphics[width=.9\linewidth]{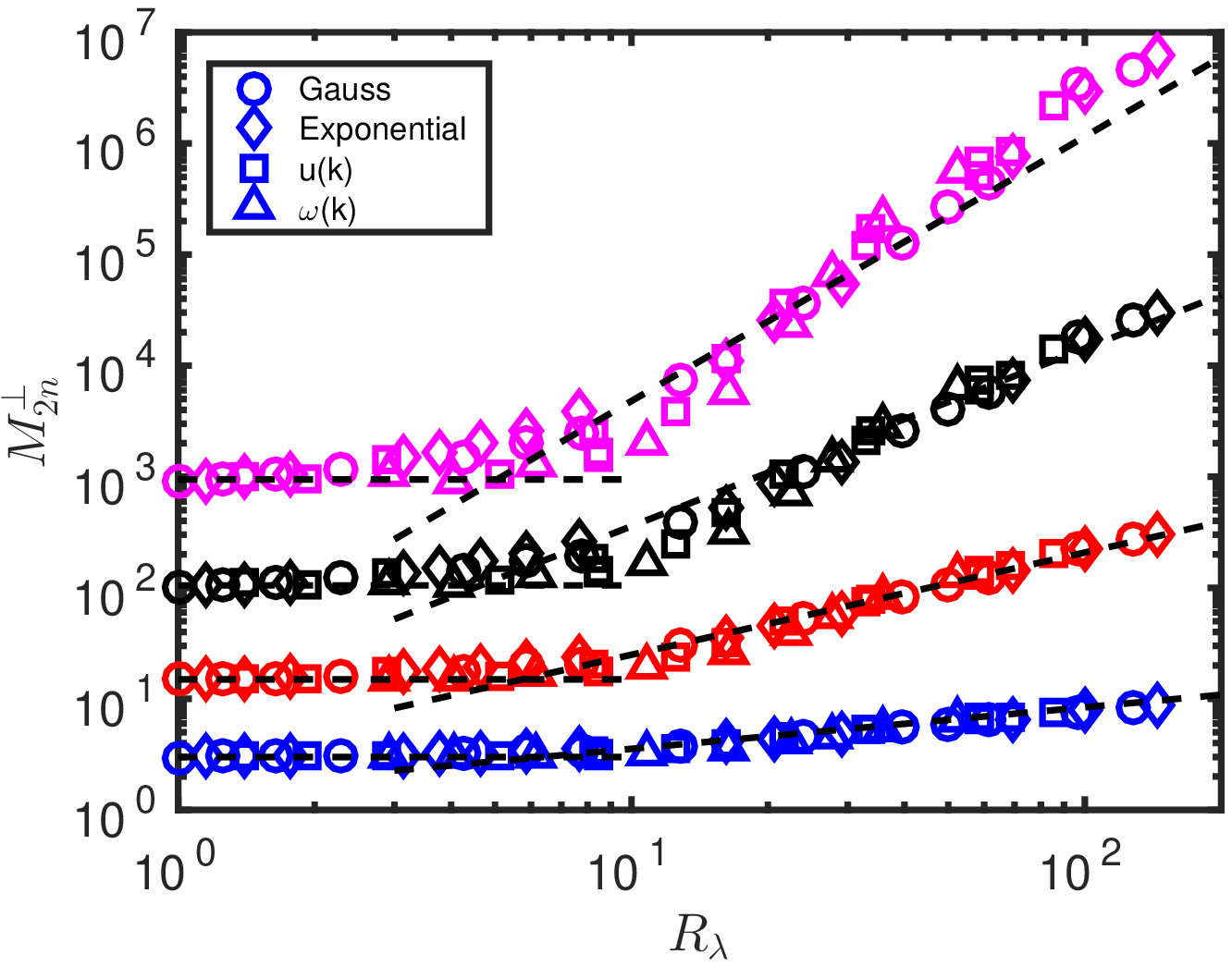}
\begin{picture}(0,0)
\put(-115,320){a)}
\put(-130,150){b)}
\end{picture}
\caption{
Moments of (a) longitudinal and (b) transverse velocity gradients for $2n=4$ (blue), $6$ (red), $8$ (black), and
$10$ (magenta). 
The horizontal dashed lines correspond to Gaussian moments, and power-laws correspond
to $\re^{2d_n}$ from \cite{YakhotDonzis2018} where the $d_n$ are given by Eq.~(\ref{dneqn}), with $\recrit=9.89$. The transverse moments in (b) follow steeper power laws than the longitudinal moments in (a).
	}
\label{vmoms}
\end{figure}

We are interested in the moments of the derivative $q$ in the form
$M_{n}^{q}={\langle q^{n}\rangle}/ \langle q\rangle^n$
where 
$n$ is the order of the moment.
The moments of longitudinal
velocity gradients from simulations with different forcing mechanisms
are shown in Fig.~\ref{vmoms}a. 
They show a composite of a 
low-$\re$ Gaussian asymptote (dashed horizontal lines) 
and a transition to anomalous scaling (dashed lines showing power-laws). 
The Gaussian asymptote for
low-$\re$, the onset of transition, and the exponents of power-law regime are all essentially independent of
forcing.
The scaling for longitudinal gradients can be accurately fitted 
by the analytical derivation of power laws in \cite{YakhotDonzis2018} (dashed lines)
assuming that the low-$\re$ moments are Gaussian and a universal transition
occurs at $\recrit=9.89$ (Eq.~(\ref{dneqn})). 
Open circles for Gaussian forcing follow the earlier result \cite{YakhotDonzis2017,YakhotDonzis2018} quite well.
Note that the power-law behavior is 
traditionally expected at much higher $\re$ 
than those found here. 

In Fig.~\ref{vmoms}b, we have plotted the moments of transverse velocity
gradients (symbols)
along with the scaling predicted from Eq.~(\ref{dneqn}) for longitudinal
gradients (dashed lines).
As for the longitudinal gradients, the transverse velocity gradient moments exhibit
a low-$\re$ Gaussian 
asymptote and transition to power-law scaling
beyond a small value of $\re$, and are also independent of the
large-scale forcing mechanism. 
However, two differences become clear when Figs.~\ref{vmoms} a and b are compared: the transition for high-moments occurs at a lower $\re$ than 10, and the moments of transverse gradients grow faster than
longitudinal gradient moments, with this tendency increasing with increasing moment order. 
This is consistent
with the claims of 
Refs.~\cite{ChenEtAl1997,Dhruva1997,GotohEtAl2002, IyerEtAl2020,gotoh2022} 
that 
transverse gradients are more intermittent than the longitudinal. 
We emphasize that, while the scaling exponents for different
gradients are different, the behavior of a given gradient is independent of forcing. 
The theory \cite{YakhotDonzis2018} allows for this possibility, so the constants in Eq.~(\ref{dneqn}) 
depend on whether they refer to the longitudinal or transverse gradients. 

To quantitatively support the observations just made,
we now compute the transition Reynolds
number as well as exponents in the power-law regime.
In the previous work \cite{YakhotDonzis2018}, only the power-law part was used for fitting purposes but this method is sensitive to the fitting
range \cite{gotoh2022}.
We mitigate this problem 
by fitting the entire data by a single
functional form for $M^q_{2n}$ that captures both the low-$\re$ asymptote and the
power-law part. Such a procedure of using scaling functions, rather than the power-law part alone, is more reliable for
obtaining scaling parameters
\cite{StolovitzkyEtAl1993}.
Since we have no analytical guidance on the full details of the transition, we can pragmatically propose
the following functional form that satisfies our requirements:
\begin{eqnarray}
M_{2n}^{q}=C_{2n}^q+\alpha_{2n}^{q}C_{2n}^{q}\left(\re\over \retr (2n)\right)^{\beta_{2n}^q}.
\end{eqnarray}
Here $\alpha_{2n}^q$ is expected to be of the order unity,
and the other three fitting parameters are
the low-$\re$ asymptote $C_{2n}^q$, the transition Reynolds
number $\retr(2n)$, and the high-$\re$ scaling exponent $\beta_{2n}^q$. 
Since $\alpha_{2n}^q$ and $C_{2n}^q$ appear as
a product, there are only three independent fitting parameters,
so 
we rewrite the above form as
\begin{eqnarray}
M_{2n}^{q}=C_{2n}^{q}\left(
1+\left(\re\over b_{2n}^{q}\right)^{\beta_{2n}^{q}}
\right),
\label{fiteqn}
\end{eqnarray}
where $b_{2n}^{q}= \retr(2n)/(\alpha_{2n}^q)^{1/\beta_{2n}^{q}}$ 
for a given quantity $q$ and order $2n$. Stable and accurate fits are possible if the 
data extend at least up to $\re = O(100)$. 

\begin{figure}
\centering
\includegraphics[trim={0.5cm 0cm 0cm 0cm},clip,width=.87\linewidth]{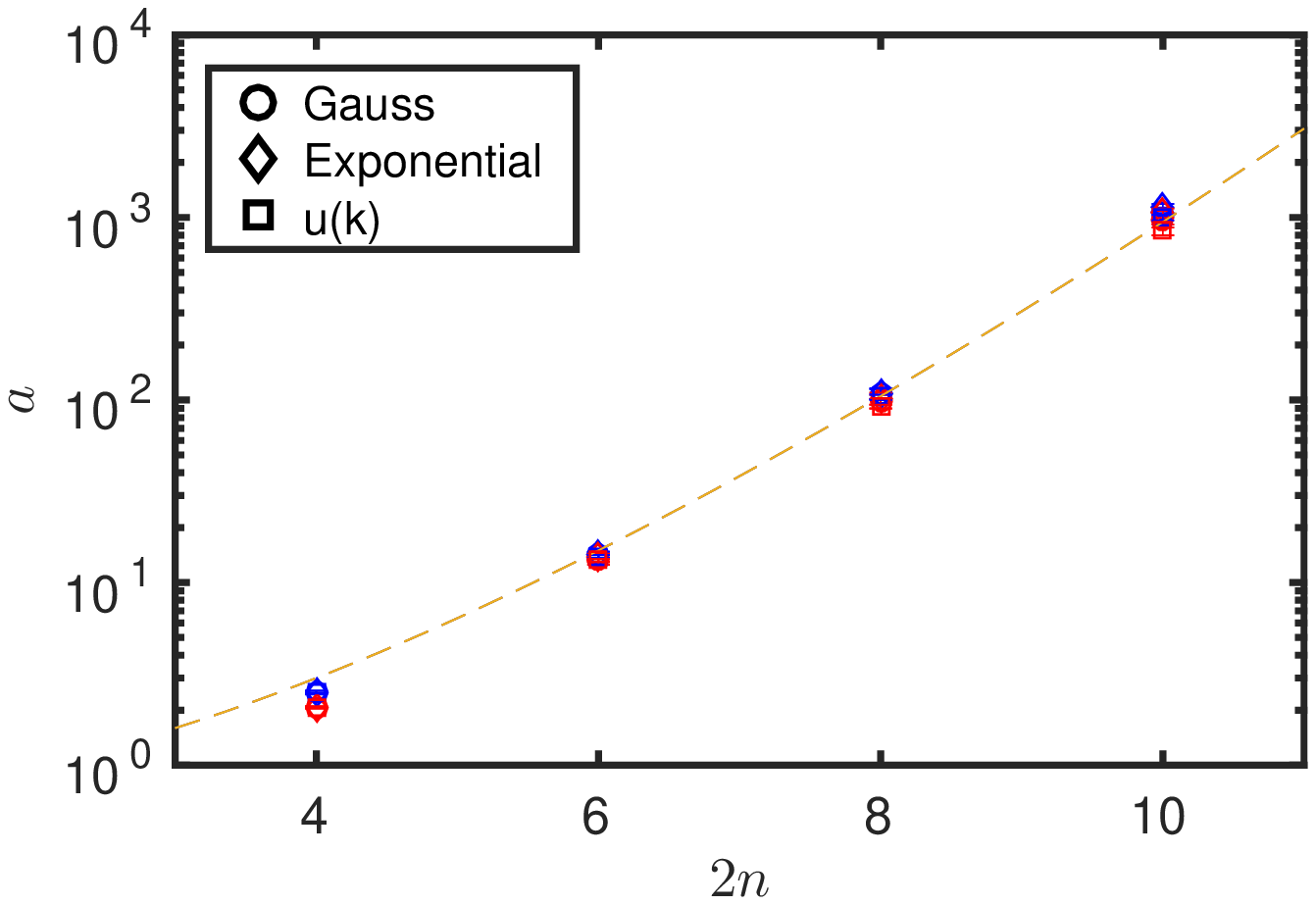}
\includegraphics[trim={1.5cm 0cm 1.5cm 0.75cm},clip,width=0.9\linewidth]{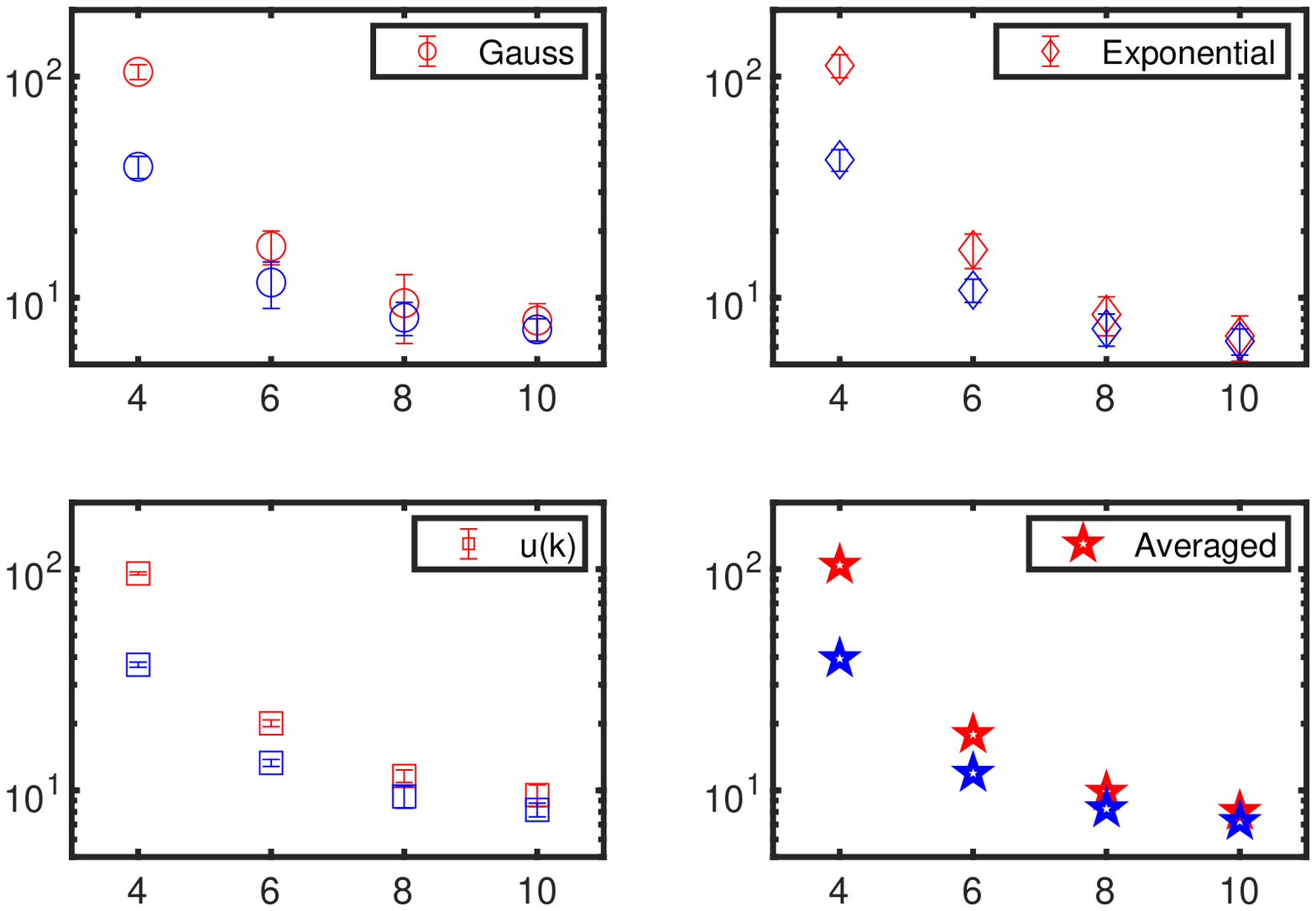}
\includegraphics[trim={0.5cm 0cm 0cm 0cm},clip,width=.88\linewidth]{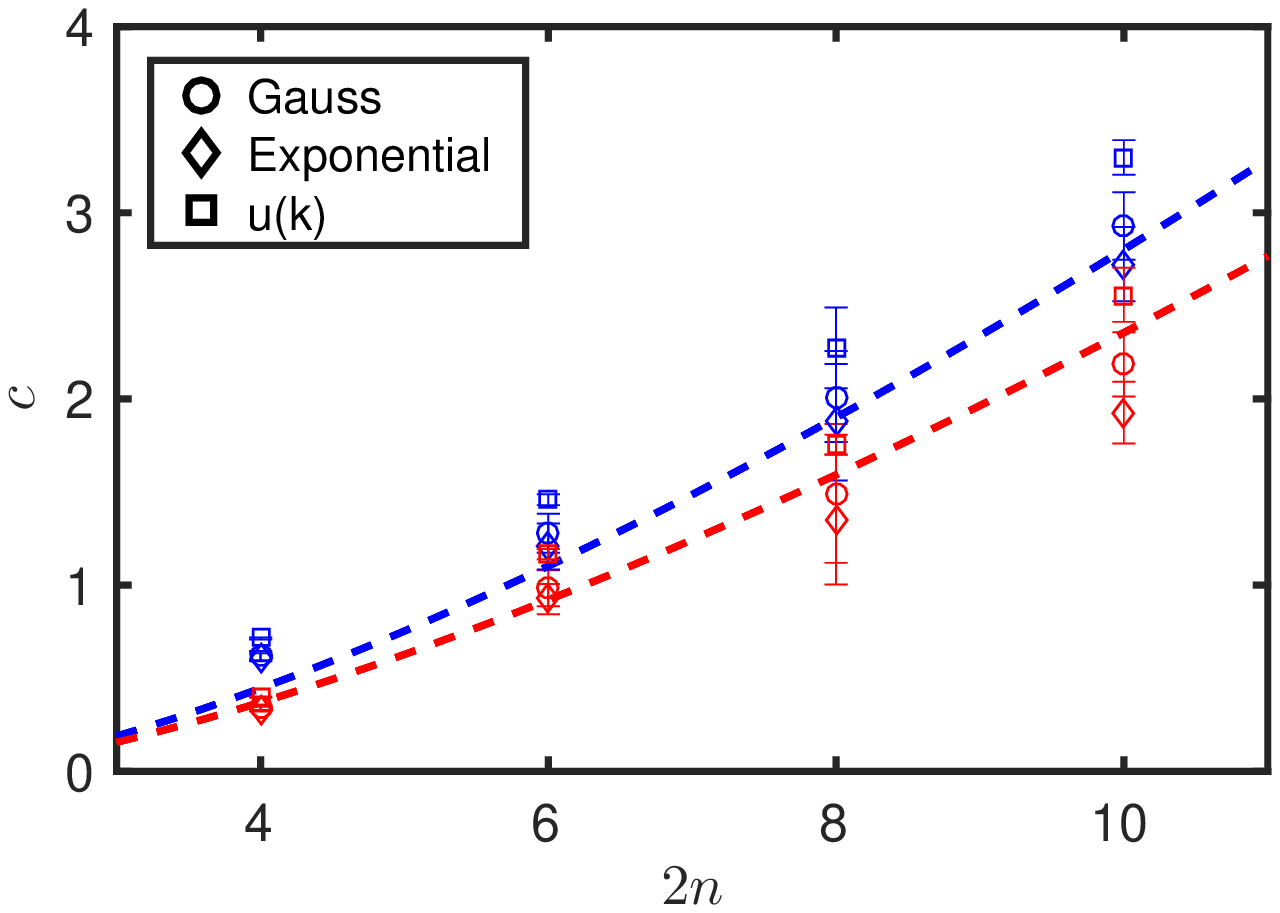}
\begin{picture}(0,0)
\put(-120,450){a)}
\put(-170,305){b)}
\put(-70,305){c)}
\put(-166,227){d)}
\put(-67,227){e)}
\put(-120,140){f)}
\put(-235,410){$C_{2n}^{q}$}
\put(-235,90){$\beta_{2n}^{q}$}
\put(-235,205){$b_{2n}^{q}$}
\put(-235,285){$b_{2n}^{q}$}
\put(-110,205){$b_{2n}^{q}$}
\put(-110,285){$b_{2n}^{q}$}

\put(-166,245){$2n$}
\put(-166,165){$2n$}
\put(-55,245){$2n$}
\put(-55,165){$2n$}
\end{picture}
\caption{
	Fit parameters for moments of longitudinal (red) and transverse (blue) 
	velocity gradients. (a) The low-$\re$ asymptotes, with the orange line denoting Gaussian moments. The constant $b_{2n}^q \propto R_{\lambda}^{tr}(n)$ are for the following forcing: (b) Gaussian (c) Exponential (d) u(k). (e) The mean values of $b_{2n}^q$ for the three forcing schemes. (f) Scaling exponents for the different forcing methods.
	The red and blue lines correspond, respectively,
to $\hat{R}_{\lambda,tr}=9.89$ and $2$ in Eq.~(\ref{dneqn}).
	We generate 50 synthetic data points from a
normal distribution with the same mean and variance as those observed for each realization 
of the DNS data.
We then perform the fit with Eq.~(\ref{fiteqn}) on these datasets
to generate a PDF for each fitting parameter. The $95\%$ confidence interval
for each parameter is generated from this PDF and shown in the figures. In many instances, the error bars are not much bigger than the symbol size. 
	}

\label{vmomsfit}
\end{figure}

The low-$\re$ asymptotes for the moments $(=C_{2n}^{q})$ so obtained are plotted in
Fig.~\ref{vmomsfit}a along with the appropriate Gaussian values (dashed line).
The two are very close to each other for both longitudinal and transverse
moments for all forcing methods, consistent with the
observation in Fig.~\ref{vmoms}. 
The parameter $b_{2n}^q$, which is proportional
to the transition $\re$,
decreases with increasing moment order, supporting the theoretical results of \cite{YakhotDonzis2017}.
For the lowest order ($n=2$), the values of 
$b_{2n}^q$ are $O(10)$.
Obtaining accurate values of $b_{2n}^q$ at low orders 
is challenging because this parameter is essentially the intersection of the low-$\re$ and high-$\re$ asymptotes which, as seen in Fig.~\ref{vmoms}, become closer to being co-linear, making the problem ill-conditioned.
For a given forcing, the transition $\re$ is lower for
transverse gradients (blue symbols) than for the longitudinal (red
symbols). Although small differences exist for $b_{2n}^{q}$ at a given $n$ for different forcing methods, they are within statistical
bounds. In Fig.~\ref{vmomsfit}e, the average values of
of $b_{2n}^q$ over all three forcing schemes clearly show the persistence of differences between
transverse and longitudinal gradients. 

The power-law exponent $\beta_{2n}^q=2d_n^q$ is plotted in
Fig.~\ref{vmomsfit}f for the three different forcing methods.
Again, the scaling exponents are larger for transverse
gradients. 
Those for longitudinal gradients are consistent with
earlier measurements made in isotropic turbulence, turbulent channel,
and Rayleigh-B\'enard convection
\cite{GotohEtAl2002,WatanabeGotoh2007,SchumacherEtAl2014,PandeySchumacher2020,gotoh2022}.
We also include Eq.~(\ref{dneqn}) (dashed line) which fit the data
with
$\recrit=9.89$ (red) and $2$ (blue) for longitudinal and transverse
gradients, respectively. Overall, the physical picture is that 
large transverse gradients acquire their high-$\re$ asymptotic 
behavior at lower Reynolds number than longitudinal ones.
The faster growth of transverse moments 
implies smaller scaling exponents of
transverse structure functions in the inertial range. 
This result has recently been reported in \cite{IyerEtAl2020} using data from
$\re\gtrsim 650$, much larger than those reported here. 

Yakhot \& Donzis \cite{YakhotDonzis2018}
allowed for different exponents for longitudinal and transverse
gradients, but
did not provide a physical reasoning. Recent work
\cite{sreenivasanyakhot2021,IyerEtAl2020} argues that high-order moments of velocity increments in the inertial range
are decreasingly affected by pressure
gradients, with two possible consequences.
First, it leads to 
stronger fluctuations and a transition at a lower-$\re$ for
high-order moments. Second, the transverse fluctuations are even less susceptible to pressure effects, possibly leading to differences in power-law scaling between longitudinal and transverse gradients.  

\begin{figure}
\centering
\vspace{0.5cm}
\includegraphics[width=0.9\linewidth]{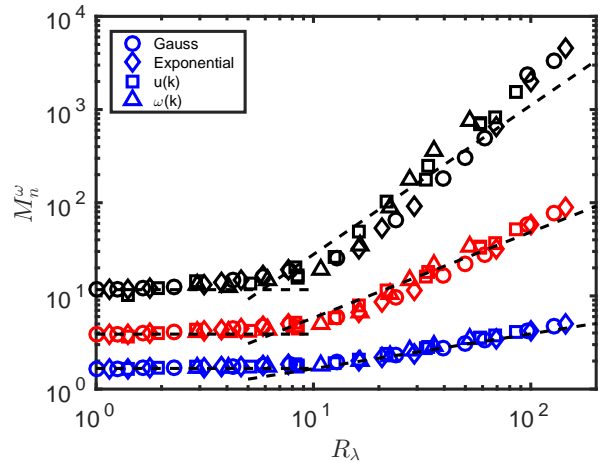}
\caption{\label{enstrophy}
Moments of enstrophy for $n=2$ (blue), $3$ (red), $4$ (black). Horizontal lines correspond to moments
of $\chi^2$ distribution with 3 degrees of freedom. The power-laws corresponds
to $\la\epsilon^n\ra/\la\epsilon\ra^n\propto\re^{d_n}$. }
\end{figure}

Velocity gradients are important, in part, because they 
combine to form two quantities of particular interest in turbulence theory---the energy dissipation rate and enstrophy density ($\Omega_i=|\boldsymbol{\omega}|^2$ 
where $\boldsymbol{\omega}$ is the vorticity vector).
While the moments of dissipation were
shown \cite{SchumacherEtAl2014,gotoh2022} to follow power-law scaling 
even at moderate $\re$, not much is known about enstrophy.
We examine it here. In Fig.~\ref{enstrophy},
we plot the moments of enstrophy (symbols) for different $\re$
and forcing schemes. The dashed power-laws
correspond to those
observed in \cite{SchumacherEtAl2014,PandeySchumacher2020,gotoh2022} 
and predicted by \cite{YakhotDonzis2017,YakhotDonzis2018}.
For $\re\lesssim 10$, 
the asymptotic values correspond to moments of the $\chi^2$ distribution
with three degrees of freedom (dashed horizontal lines).
The agreement towards the low-$\re$ asymptote is expected as enstrophy is the sum of squares of three
transverse gradients, each of which is Gaussian 
and independent of the other two, given the weak coupling expected
at this low Reynolds numbers.
Similarly dissipation exhibits $\chi^2$ statistics with five degrees
of freedom as the incompressibility 
condition constrains only five gradients
to be independent.
This feature is indeed observed in our data as well (not shown here).
Similar to individual gradients,
the transition to the anomalous regime for enstrophy appears to be independent of the details of forcing. We also note that scaling exponents for enstrophy are 
larger than those for dissipation (shown as dashed-line power laws
in \rfig{enstrophy}).
Enstrophy moments grow faster than those of dissipation, 
increasingly so at higher orders.
These observations are consistent with 
the available evidence at much higher $\re$ that extreme events in enstrophy are 
more probable than in dissipation \cite{YeungEtAl2018,ChenEtAl1997,gotoh2022}. 
We thus conclude that high-$\re$ behaviors for dissipation and 
enstrophy are also incipient at low Reynolds numbers $\re\sim O(10)$, which marks the transition.

\section{Conclusions}
We have shown here that anomalous scaling 
for velocity gradients and enstrophy emerges 
at $\re\sim O(10)$, much lower
than traditionally expected, consistent with \cite{YakhotDonzis2017}.
Using different driving mechanisms at large scales,
we have further shown that this scaling behavior is independent of the details of forcing.
Moments of longitudinal
and transverse velocity gradients, and those of dissipation and enstrophy, possess different sets of scaling exponents.
In particular, the
scaling exponents are larger for transverse gradients, 
consistent with the literature \cite{IyerEtAl2020,YeungEtAl2015,Dhruva1997,YeungEtAl2018}.
All scaling exponents can be predicted by the theory \cite{YakhotDonzis2017} by knowing the transition $\re$.
In particular, the theory predicts that higher exponents
will be obtained if the transition occurs at a lower $\re$. 
This is indeed what we observe. 

Another interesting point is that the theory \cite{YakhotDonzis2018} relates velocity gradient exponents to those of structure functions in the inertial range. 
Note that all results here are 
for $1\lesssim \re\lesssim 100$, which are lower 
than those needed for an inertial range to emerge \cite{DS2010b}. 
Yet, the inertial range exponents calculated 
from the exponents $\beta_{2n}^q$ obtained here, using the theory,
are close to those observed 
in simulations and experiments at high $\re$ where an inertial range does exist.
A potential implication
is that {\it certain} high Reynolds features can be studied using data from well resolved 
DNS at low to moderate $\re$, and do not need very high $\re$. 
From a physical point of view, the inertial 
range anomalies are the result of intermittency at small scales
which appear at low $\re$ even without an inertial range. 
In this view, the inertial range 
emerges only as an intermediate constraint to 
match the Gaussian large scales with the anomalous dissipative scales. 

Finally, we have evidence to support the present view in passive scalar
advection and compressible turbulence
\cite{khurshidthesis,gotoh2022}---also for the Burgers equation that is studied, e.g., in Ref.~\cite{FriedrichEtAl2018}. These
results will be reported elsewhere.

{\it Acknowledgments:} The authors gratefully acknowledge many insightful discussions with Victor Yakhot. SK was partially supported by the
National Science Foundation Grant\# 2127309 to the Computing Research Association for the CIFellows 2021 Project. 
\bibliography{main}

\begin{thebibliography}{46}%
\makeatletter
\providecommand \@ifxundefined [1]{%
 \@ifx{#1\undefined}
}%
\providecommand \@ifnum [1]{%
 \ifnum #1\expandafter \@firstoftwo
 \else \expandafter \@secondoftwo
 \fi
}%
\providecommand \@ifx [1]{%
 \ifx #1\expandafter \@firstoftwo
 \else \expandafter \@secondoftwo
 \fi
}%
\providecommand \natexlab [1]{#1}%
\providecommand \enquote  [1]{``#1''}%
\providecommand \bibnamefont  [1]{#1}%
\providecommand \bibfnamefont [1]{#1}%
\providecommand \citenamefont [1]{#1}%
\providecommand \href@noop [0]{\@secondoftwo}%
\providecommand \href [0]{\begingroup \@sanitize@url \@href}%
\providecommand \@href[1]{\@@startlink{#1}\@@href}%
\providecommand \@@href[1]{\endgroup#1\@@endlink}%
\providecommand \@sanitize@url [0]{\catcode `\\12\catcode `\$12\catcode
  `\&12\catcode `\#12\catcode `\^12\catcode `\_12\catcode `\%12\relax}%
\providecommand \@@startlink[1]{}%
\providecommand \@@endlink[0]{}%
\providecommand \url  [0]{\begingroup\@sanitize@url \@url }%
\providecommand \@url [1]{\endgroup\@href {#1}{\urlprefix }}%
\providecommand \urlprefix  [0]{URL }%
\providecommand \Eprint [0]{\href }%
\providecommand \doibase [0]{https://doi.org/}%
\providecommand \selectlanguage [0]{\@gobble}%
\providecommand \bibinfo  [0]{\@secondoftwo}%
\providecommand \bibfield  [0]{\@secondoftwo}%
\providecommand \translation [1]{[#1]}%
\providecommand \BibitemOpen [0]{}%
\providecommand \bibitemStop [0]{}%
\providecommand \bibitemNoStop [0]{.\EOS\space}%
\providecommand \EOS [0]{\spacefactor3000\relax}%
\providecommand \BibitemShut  [1]{\csname bibitem#1\endcsname}%
\let\auto@bib@innerbib\@empty
\bibitem [{\citenamefont {Kolmogorov}(1941)}]{K41}%
  \BibitemOpen
  \bibfield  {author} {\bibinfo {author} {\bibfnamefont {A.~N.}\ \bibnamefont
  {Kolmogorov}},\ }\bibfield  {title} {\bibinfo {title} {Local structure of
  turbulence in an incompressible fluid for very large {R}eynolds numbers},\
  }\href@noop {} {\bibfield  {journal} {\bibinfo  {journal} {Dokl. Akad. Nauk.
  SSSR}\ }\textbf {\bibinfo {volume} {30}},\ \bibinfo {pages} {299} (\bibinfo
  {year} {1941})}\BibitemShut {NoStop}%
\bibitem [{\citenamefont {Monin}\ and\ \citenamefont {Yaglom}(1975)}]{MY.II}%
  \BibitemOpen
  \bibfield  {author} {\bibinfo {author} {\bibfnamefont {A.~S.}\ \bibnamefont
  {Monin}}\ and\ \bibinfo {author} {\bibfnamefont {A.~M.}\ \bibnamefont
  {Yaglom}},\ }\href@noop {} {\emph {\bibinfo {title} {Statistical Fluid
  Mechanics}}},\ Vol.~\bibinfo {volume} {2}\ (\bibinfo  {publisher} {MIT
  Press},\ \bibinfo {year} {1975})\BibitemShut {NoStop}%
\bibitem [{\citenamefont {Batchelor}(1953)}]{batchelor1953}%
  \BibitemOpen
  \bibfield  {author} {\bibinfo {author} {\bibfnamefont {G.~K.}\ \bibnamefont
  {Batchelor}},\ }\href@noop {} {\emph {\bibinfo {title} {The {t}heory of
  {h}omogeneous {t}urbulence}}}\ (\bibinfo  {publisher} {Cambridge University
  Press},\ \bibinfo {year} {1953})\BibitemShut {NoStop}%
\bibitem [{\citenamefont {Kolmogorov}(1962)}]{K62}%
  \BibitemOpen
  \bibfield  {author} {\bibinfo {author} {\bibfnamefont {A.~N.}\ \bibnamefont
  {Kolmogorov}},\ }\bibfield  {title} {\bibinfo {title} {A refinement of
  previous hypotheses concerning the local structure of turbulence in a viscous
  incompressible fluid at high {R}eynolds number},\ }\href@noop {} {\bibfield
  {journal} {\bibinfo  {journal} {J. Fluid Mech.}\ }\textbf {\bibinfo {volume}
  {13}},\ \bibinfo {pages} {82} (\bibinfo {year} {1962})}\BibitemShut {NoStop}%
\bibitem [{\citenamefont {Frisch}(1995)}]{frisch95}%
  \BibitemOpen
  \bibfield  {author} {\bibinfo {author} {\bibfnamefont {U.}~\bibnamefont
  {Frisch}},\ }\href@noop {} {\emph {\bibinfo {title} {Turbulence}}}\ (\bibinfo
   {publisher} {Cambridge University Press},\ \bibinfo {year}
  {1995})\BibitemShut {NoStop}%
\bibitem [{\citenamefont {Sreenivasan}\ and\ \citenamefont
  {Antonia}(1997)}]{SA97}%
  \BibitemOpen
  \bibfield  {author} {\bibinfo {author} {\bibfnamefont {K.~R.}\ \bibnamefont
  {Sreenivasan}}\ and\ \bibinfo {author} {\bibfnamefont {R.~A.}\ \bibnamefont
  {Antonia}},\ }\bibfield  {title} {\bibinfo {title} {The phenomenology of
  small-scale turbulence},\ }\href@noop {} {\bibfield  {journal} {\bibinfo
  {journal} {Annu. Rev. Fluid Mech.}\ }\textbf {\bibinfo {volume} {29}},\
  \bibinfo {pages} {435} (\bibinfo {year} {1997})}\BibitemShut {NoStop}%
\bibitem [{\citenamefont {She}\ and\ \citenamefont
  {Leveque}(1994)}]{SheLeveque1994}%
  \BibitemOpen
  \bibfield  {author} {\bibinfo {author} {\bibfnamefont {Z.-S.}\ \bibnamefont
  {She}}\ and\ \bibinfo {author} {\bibfnamefont {E.}~\bibnamefont {Leveque}},\
  }\bibfield  {title} {\bibinfo {title} {Universal scaling laws in fully
  developed turbulence},\ }\href@noop {} {\bibfield  {journal} {\bibinfo
  {journal} {Phys. Rev. Lett.}\ }\textbf {\bibinfo {volume} {72}},\ \bibinfo
  {pages} {336} (\bibinfo {year} {1994})}\BibitemShut {NoStop}%
\bibitem [{\citenamefont {Sreenivasan}\ and\ \citenamefont
  {Yakhot}(2021)}]{sreenivasanyakhot2021}%
  \BibitemOpen
  \bibfield  {author} {\bibinfo {author} {\bibfnamefont {K.~R.}\ \bibnamefont
  {Sreenivasan}}\ and\ \bibinfo {author} {\bibfnamefont {V.}~\bibnamefont
  {Yakhot}},\ }\bibfield  {title} {\bibinfo {title} {Dynamics of
  three-dimensional turbulence from {N}avier-{S}tokes equations},\ }\href@noop
  {} {\bibfield  {journal} {\bibinfo  {journal} {Phys. Rev. Fluids}\ }\textbf
  {\bibinfo {volume} {6}},\ \bibinfo {pages} {104604} (\bibinfo {year}
  {2021})}\BibitemShut {NoStop}%
\bibitem [{\citenamefont {Yakhot}\ \emph {et~al.}(1992)\citenamefont {Yakhot},
  \citenamefont {Orszag}, \citenamefont {Thangam}, \citenamefont {Gatski},\
  and\ \citenamefont {Speziale}}]{yakhot1992}%
  \BibitemOpen
  \bibfield  {author} {\bibinfo {author} {\bibfnamefont {V.}~\bibnamefont
  {Yakhot}}, \bibinfo {author} {\bibfnamefont {S.}~\bibnamefont {Orszag}},
  \bibinfo {author} {\bibfnamefont {S.}~\bibnamefont {Thangam}}, \bibinfo
  {author} {\bibfnamefont {T.}~\bibnamefont {Gatski}},\ and\ \bibinfo {author}
  {\bibfnamefont {C.}~\bibnamefont {Speziale}},\ }\bibfield  {title} {\bibinfo
  {title} {Development of turbulence models for shear flows by a double
  expansion technique},\ }\href@noop {} {\bibfield  {journal} {\bibinfo
  {journal} {Phys. Fluids A: Fluid Dyn.}\ }\textbf {\bibinfo {volume} {4}},\
  \bibinfo {pages} {1510} (\bibinfo {year} {1992})}\BibitemShut {NoStop}%
\bibitem [{\citenamefont {Benzi}\ \emph {et~al.}(1993)\citenamefont {Benzi},
  \citenamefont {Ciliberto}, \citenamefont {Tripiccione}, \citenamefont
  {Baudet}, \citenamefont {Massaioli},\ and\ \citenamefont
  {Succi}}]{BenziEtAl1993}%
  \BibitemOpen
  \bibfield  {author} {\bibinfo {author} {\bibfnamefont {R.}~\bibnamefont
  {Benzi}}, \bibinfo {author} {\bibfnamefont {S.}~\bibnamefont {Ciliberto}},
  \bibinfo {author} {\bibfnamefont {R.}~\bibnamefont {Tripiccione}}, \bibinfo
  {author} {\bibfnamefont {C.}~\bibnamefont {Baudet}}, \bibinfo {author}
  {\bibfnamefont {F.}~\bibnamefont {Massaioli}},\ and\ \bibinfo {author}
  {\bibfnamefont {S.}~\bibnamefont {Succi}},\ }\bibfield  {title} {\bibinfo
  {title} {Extended self-similarity in turbulent flows},\ }\href@noop {}
  {\bibfield  {journal} {\bibinfo  {journal} {Phys. Rev. E}\ }\textbf {\bibinfo
  {volume} {48}} (\bibinfo {year} {1993})}\BibitemShut {NoStop}%
\bibitem [{\citenamefont {Benzi}\ \emph {et~al.}(1995)\citenamefont {Benzi},
  \citenamefont {Ciliberto}, \citenamefont {Baudet},\ and\ \citenamefont
  {Chavarria}}]{BenziEtAl1995}%
  \BibitemOpen
  \bibfield  {author} {\bibinfo {author} {\bibfnamefont {R.}~\bibnamefont
  {Benzi}}, \bibinfo {author} {\bibfnamefont {S.}~\bibnamefont {Ciliberto}},
  \bibinfo {author} {\bibfnamefont {C.}~\bibnamefont {Baudet}},\ and\ \bibinfo
  {author} {\bibfnamefont {G.}~\bibnamefont {Chavarria}},\ }\bibfield  {title}
  {\bibinfo {title} {On the scaling of three-dimensional homogeneous and
  isotropic turbulence},\ }\href@noop {} {\bibfield  {journal} {\bibinfo
  {journal} {Phys. D}\ }\textbf {\bibinfo {volume} {80}},\ \bibinfo {pages}
  {385} (\bibinfo {year} {1995})}\BibitemShut {NoStop}%
\bibitem [{\citenamefont {Iyer}\ \emph {et~al.}(2017)\citenamefont {Iyer},
  \citenamefont {Sreenivasan},\ and\ \citenamefont {Yeung}}]{IyerEtAl2017}%
  \BibitemOpen
  \bibfield  {author} {\bibinfo {author} {\bibfnamefont {K.~P.}\ \bibnamefont
  {Iyer}}, \bibinfo {author} {\bibfnamefont {K.~R.}\ \bibnamefont
  {Sreenivasan}},\ and\ \bibinfo {author} {\bibfnamefont {P.~K.}\ \bibnamefont
  {Yeung}},\ }\bibfield  {title} {\bibinfo {title} {Reynolds number scaling of
  velocity increments in isotropic turbulence},\ }\href@noop {} {\bibfield
  {journal} {\bibinfo  {journal} {Phys. Rev. E}\ }\textbf {\bibinfo {volume}
  {95}} (\bibinfo {year} {2017})}\BibitemShut {NoStop}%
\bibitem [{\citenamefont {Iyer}\ \emph {et~al.}(2020)\citenamefont {Iyer},
  \citenamefont {Sreenivasan},\ and\ \citenamefont {Yeung}}]{IyerEtAl2020}%
  \BibitemOpen
  \bibfield  {author} {\bibinfo {author} {\bibfnamefont {K.~P.}\ \bibnamefont
  {Iyer}}, \bibinfo {author} {\bibfnamefont {K.~R.}\ \bibnamefont
  {Sreenivasan}},\ and\ \bibinfo {author} {\bibfnamefont {P.~K.}\ \bibnamefont
  {Yeung}},\ }\bibfield  {title} {\bibinfo {title} {Scaling exponents saturate
  in three-dimensional isotropic turbulence},\ }\href@noop {} {\bibfield
  {journal} {\bibinfo  {journal} {Phys. Rev. Fluids}\ }\textbf {\bibinfo
  {volume} {5}},\ \bibinfo {pages} {054605} (\bibinfo {year}
  {2020})}\BibitemShut {NoStop}%
\bibitem [{\citenamefont {Tsuji}(2004)}]{tsuji2004}%
  \BibitemOpen
  \bibfield  {author} {\bibinfo {author} {\bibfnamefont {Y.}~\bibnamefont
  {Tsuji}},\ }\bibfield  {title} {\bibinfo {title} {Intermittency effect on
  energy spectrum in high-reynolds number turbulence},\ }\href@noop {}
  {\bibfield  {journal} {\bibinfo  {journal} {Phys. Fluids}\ }\textbf {\bibinfo
  {volume} {16}} (\bibinfo {year} {2004})}\BibitemShut {NoStop}%
\bibitem [{\citenamefont {Ishihara}\ \emph {et~al.}(2009)\citenamefont
  {Ishihara}, \citenamefont {Gotoh},\ and\ \citenamefont {Kaneda}}]{IGK2009}%
  \BibitemOpen
  \bibfield  {author} {\bibinfo {author} {\bibfnamefont {T.}~\bibnamefont
  {Ishihara}}, \bibinfo {author} {\bibfnamefont {T.}~\bibnamefont {Gotoh}},\
  and\ \bibinfo {author} {\bibfnamefont {Y.}~\bibnamefont {Kaneda}},\
  }\bibfield  {title} {\bibinfo {title} {Study of high-{R}eynolds number
  isotropic turbulence by direct numerical simulation},\ }\href@noop {}
  {\bibfield  {journal} {\bibinfo  {journal} {Annu. Rev. Fluid Mech.}\ }\textbf
  {\bibinfo {volume} {41}},\ \bibinfo {pages} {165} (\bibinfo {year}
  {2009})}\BibitemShut {NoStop}%
\bibitem [{\citenamefont {Donzis}\ and\ \citenamefont
  {Sreenivasan}(2010)}]{DS2010b}%
  \BibitemOpen
  \bibfield  {author} {\bibinfo {author} {\bibfnamefont {D.~A.}\ \bibnamefont
  {Donzis}}\ and\ \bibinfo {author} {\bibfnamefont {K.~R.}\ \bibnamefont
  {Sreenivasan}},\ }\bibfield  {title} {\bibinfo {title} {The bottleneck effect
  and the {K}olmogorov constant in isotropic turbulence},\ }\href@noop {}
  {\bibfield  {journal} {\bibinfo  {journal} {J. Fluid Mech.}\ }\textbf
  {\bibinfo {volume} {657}},\ \bibinfo {pages} {171} (\bibinfo {year}
  {2010})}\BibitemShut {NoStop}%
\bibitem [{\citenamefont {Corrsin}(1958)}]{Corrsin1958}%
  \BibitemOpen
  \bibfield  {author} {\bibinfo {author} {\bibfnamefont {S.}~\bibnamefont
  {Corrsin}},\ }\bibfield  {title} {\bibinfo {title} {On local isotropy in
  turbulent shear flows},\ }\href@noop {} {\bibfield  {journal} {\bibinfo
  {journal} {NACA R \& M}\ }\textbf {\bibinfo {volume} {58B11}} (\bibinfo
  {year} {1958})}\BibitemShut {NoStop}%
\bibitem [{\citenamefont {Schumacher}\ \emph {et~al.}(2014)\citenamefont
  {Schumacher}, \citenamefont {Scheel}, \citenamefont {Krasnov}, \citenamefont
  {Donzis}, \citenamefont {Yakhot},\ and\ \citenamefont
  {Sreenivasan}}]{SchumacherEtAl2014}%
  \BibitemOpen
  \bibfield  {author} {\bibinfo {author} {\bibfnamefont {J.}~\bibnamefont
  {Schumacher}}, \bibinfo {author} {\bibfnamefont {J.~D.}\ \bibnamefont
  {Scheel}}, \bibinfo {author} {\bibfnamefont {D.}~\bibnamefont {Krasnov}},
  \bibinfo {author} {\bibfnamefont {D.~A.}\ \bibnamefont {Donzis}}, \bibinfo
  {author} {\bibfnamefont {V.}~\bibnamefont {Yakhot}},\ and\ \bibinfo {author}
  {\bibfnamefont {K.~R.}\ \bibnamefont {Sreenivasan}},\ }\bibfield  {title}
  {\bibinfo {title} {Small-scale universality in fluid turbulence},\
  }\href@noop {} {\bibfield  {journal} {\bibinfo  {journal} {Proc. Nat. Acad.
  Sci.}\ }\textbf {\bibinfo {volume} {111}},\ \bibinfo {pages} {10961}
  (\bibinfo {year} {2014})}\BibitemShut {NoStop}%
\bibitem [{\citenamefont {Yakhot}\ and\ \citenamefont
  {Donzis}(2017)}]{YakhotDonzis2017}%
  \BibitemOpen
  \bibfield  {author} {\bibinfo {author} {\bibfnamefont {V.}~\bibnamefont
  {Yakhot}}\ and\ \bibinfo {author} {\bibfnamefont {D.~A.}\ \bibnamefont
  {Donzis}},\ }\bibfield  {title} {\bibinfo {title} {Emergence of multiscaling
  in a random-force stirred fluid},\ }\href@noop {} {\bibfield  {journal}
  {\bibinfo  {journal} {Phys. Rev. Lett.}\ }\textbf {\bibinfo {volume} {119}}
  (\bibinfo {year} {2017})}\BibitemShut {NoStop}%
\bibitem [{\citenamefont {Yakhot}\ and\ \citenamefont
  {Donzis}(2018)}]{YakhotDonzis2018}%
  \BibitemOpen
  \bibfield  {author} {\bibinfo {author} {\bibfnamefont {V.}~\bibnamefont
  {Yakhot}}\ and\ \bibinfo {author} {\bibfnamefont {D.~A.}\ \bibnamefont
  {Donzis}},\ }\bibfield  {title} {\bibinfo {title} {Anomalous exponents in
  strong turbulence},\ }\href@noop {} {\bibfield  {journal} {\bibinfo
  {journal} {Phys. D: Nonlin. Phen.}\ }\textbf {\bibinfo {volume} {384}},\
  \bibinfo {pages} {12} (\bibinfo {year} {2018})}\BibitemShut {NoStop}%
\bibitem [{\citenamefont {Schumacher}\ \emph {et~al.}(2007)\citenamefont
  {Schumacher}, \citenamefont {Sreenivasan},\ and\ \citenamefont
  {Yakhot}}]{SchumacherEtAl2007a}%
  \BibitemOpen
  \bibfield  {author} {\bibinfo {author} {\bibfnamefont {J.}~\bibnamefont
  {Schumacher}}, \bibinfo {author} {\bibfnamefont {K.~R.}\ \bibnamefont
  {Sreenivasan}},\ and\ \bibinfo {author} {\bibfnamefont {V.}~\bibnamefont
  {Yakhot}},\ }\bibfield  {title} {\bibinfo {title} {Asymptotic exponents from
  low-{R}eynolds-number flows},\ }\href@noop {} {\bibfield  {journal} {\bibinfo
   {journal} {New J. Phys.}\ }\textbf {\bibinfo {volume} {9}} (\bibinfo {year}
  {2007})}\BibitemShut {NoStop}%
\bibitem [{\citenamefont {Yakhot}\ and\ \citenamefont
  {Sreenivasan}(2005)}]{YakhotSreenivasan2005}%
  \BibitemOpen
  \bibfield  {author} {\bibinfo {author} {\bibfnamefont {V.}~\bibnamefont
  {Yakhot}}\ and\ \bibinfo {author} {\bibfnamefont {K.~R.}\ \bibnamefont
  {Sreenivasan}},\ }\bibfield  {title} {\bibinfo {title} {Anomalous scaling of
  structure functions and dynamic constraints on turbulence simulations},\
  }\href@noop {} {\bibfield  {journal} {\bibinfo  {journal} {J. Stat. Phys.}\
  }\textbf {\bibinfo {volume} {121}},\ \bibinfo {pages} {823} (\bibinfo {year}
  {2005})}\BibitemShut {NoStop}%
\bibitem [{\citenamefont {Chen}\ \emph {et~al.}(1997)\citenamefont {Chen},
  \citenamefont {Sreenivasan}, \citenamefont {Nelkin},\ and\ \citenamefont
  {Cao}}]{ChenEtAl1997}%
  \BibitemOpen
  \bibfield  {author} {\bibinfo {author} {\bibfnamefont {S.}~\bibnamefont
  {Chen}}, \bibinfo {author} {\bibfnamefont {K.~R.}\ \bibnamefont
  {Sreenivasan}}, \bibinfo {author} {\bibfnamefont {M.}~\bibnamefont
  {Nelkin}},\ and\ \bibinfo {author} {\bibfnamefont {N.}~\bibnamefont {Cao}},\
  }\bibfield  {title} {\bibinfo {title} {Refined similarity hypothesis for
  transverse structure functions in fluid turbulence},\ }\href@noop {}
  {\bibfield  {journal} {\bibinfo  {journal} {Phys. Rev. Lett.}\ }\textbf
  {\bibinfo {volume} {79}},\ \bibinfo {pages} {2253} (\bibinfo {year}
  {1997})}\BibitemShut {NoStop}%
\bibitem [{\citenamefont {Donzis}\ \emph {et~al.}(2008)\citenamefont {Donzis},
  \citenamefont {Yeung},\ and\ \citenamefont {Sreenivasan}}]{DonzisEtAl2008}%
  \BibitemOpen
  \bibfield  {author} {\bibinfo {author} {\bibfnamefont {D.~A.}\ \bibnamefont
  {Donzis}}, \bibinfo {author} {\bibfnamefont {P.~K.}\ \bibnamefont {Yeung}},\
  and\ \bibinfo {author} {\bibfnamefont {K.~R.}\ \bibnamefont {Sreenivasan}},\
  }\bibfield  {title} {\bibinfo {title} {Dissipation and enstrophy in isotropic
  turbulence: Resolution effects and scaling in direct numerical simulations},\
  }\href@noop {} {\bibfield  {journal} {\bibinfo  {journal} {Phys. Fluids}\
  }\textbf {\bibinfo {volume} {20}},\ \bibinfo {pages} {45108} (\bibinfo {year}
  {2008})}\BibitemShut {NoStop}%
\bibitem [{\citenamefont {Gotoh}\ \emph {et~al.}(2002)\citenamefont {Gotoh},
  \citenamefont {Fukayama},\ and\ \citenamefont {Nakano}}]{GotohEtAl2002}%
  \BibitemOpen
  \bibfield  {author} {\bibinfo {author} {\bibfnamefont {T.}~\bibnamefont
  {Gotoh}}, \bibinfo {author} {\bibfnamefont {D.}~\bibnamefont {Fukayama}},\
  and\ \bibinfo {author} {\bibfnamefont {T.}~\bibnamefont {Nakano}},\
  }\bibfield  {title} {\bibinfo {title} {Velocity field statistics in
  homogeneous steady turbulence obtained using a high-resolution direct
  numerical simulation},\ }\href@noop {} {\bibfield  {journal} {\bibinfo
  {journal} {Phys. Fluids}\ }\textbf {\bibinfo {volume} {14}},\ \bibinfo
  {pages} {1065} (\bibinfo {year} {2002})}\BibitemShut {NoStop}%
\bibitem [{\citenamefont {Gotoh}\ and\ \citenamefont {Yang}(2022)}]{gotoh2022}%
  \BibitemOpen
  \bibfield  {author} {\bibinfo {author} {\bibfnamefont {T.}~\bibnamefont
  {Gotoh}}\ and\ \bibinfo {author} {\bibfnamefont {J.}~\bibnamefont {Yang}},\
  }\bibfield  {title} {\bibinfo {title} {Transition of fluctuations from
  gaussian state to turbulent state},\ }\href@noop {} {\bibfield  {journal}
  {\bibinfo  {journal} {Phil. Trans. Royal Soc. A}\ }\textbf {\bibinfo {volume}
  {380}},\ \bibinfo {pages} {20210097} (\bibinfo {year} {2022})}\BibitemShut
  {NoStop}%
\bibitem [{\citenamefont {Gibbon}\ and\ \citenamefont {Doering}(2005)}]{GD05}%
  \BibitemOpen
  \bibfield  {author} {\bibinfo {author} {\bibfnamefont {J.~D.}\ \bibnamefont
  {Gibbon}}\ and\ \bibinfo {author} {\bibfnamefont {C.~R.}\ \bibnamefont
  {Doering}},\ }\bibfield  {title} {\bibinfo {title} {Intermittency and
  regularity issues in 3d {N}avier-{S}tokes turbulence},\ }\href@noop {}
  {\bibfield  {journal} {\bibinfo  {journal} {Arch. Rat. Mech. \& Anal.}\
  }\textbf {\bibinfo {volume} {177}},\ \bibinfo {pages} {115} (\bibinfo {year}
  {2005})}\BibitemShut {NoStop}%
\bibitem [{\citenamefont {Beale}\ \emph {et~al.}(1984)\citenamefont {Beale},
  \citenamefont {Kato},\ and\ \citenamefont {Majda}}]{beale1984}%
  \BibitemOpen
  \bibfield  {author} {\bibinfo {author} {\bibfnamefont {J.~T.}\ \bibnamefont
  {Beale}}, \bibinfo {author} {\bibfnamefont {T.}~\bibnamefont {Kato}},\ and\
  \bibinfo {author} {\bibfnamefont {A.}~\bibnamefont {Majda}},\ }\bibfield
  {title} {\bibinfo {title} {Remarks on the breakdown of smooth solutions for
  the 3-{D} {E}uler equations},\ }\href@noop {} {\bibfield  {journal} {\bibinfo
   {journal} {Comm. Math. Phys.}\ }\textbf {\bibinfo {volume} {94}},\ \bibinfo
  {pages} {61} (\bibinfo {year} {1984})}\BibitemShut {NoStop}%
\bibitem [{\citenamefont {Yeung}\ \emph {et~al.}(2012)\citenamefont {Yeung},
  \citenamefont {Donzis},\ and\ \citenamefont {Sreenivasan}}]{YeungEtAl2012}%
  \BibitemOpen
  \bibfield  {author} {\bibinfo {author} {\bibfnamefont {P.~K.}\ \bibnamefont
  {Yeung}}, \bibinfo {author} {\bibfnamefont {D.~A.}\ \bibnamefont {Donzis}},\
  and\ \bibinfo {author} {\bibfnamefont {K.~R.}\ \bibnamefont {Sreenivasan}},\
  }\bibfield  {title} {\bibinfo {title} {Dissipation, enstrophy and pressure
  statistics in turbulence simulations at high {R}eynolds numbers},\
  }\href@noop {} {\bibfield  {journal} {\bibinfo  {journal} {J. Fluid Mech.}\
  }\textbf {\bibinfo {volume} {700}},\ \bibinfo {pages} {5} (\bibinfo {year}
  {2012})}\BibitemShut {NoStop}%
\bibitem [{\citenamefont {Schumacher}(2007)}]{SchumacherEtAl2007}%
  \BibitemOpen
  \bibfield  {author} {\bibinfo {author} {\bibfnamefont {J.}~\bibnamefont
  {Schumacher}},\ }\bibfield  {title} {\bibinfo {title} {Sub-{K}olmogorov-scale
  fluctuations in fluid turbulence},\ }\href@noop {} {\bibfield  {journal}
  {\bibinfo  {journal} {Euro. Phys. Lett.}\ }\textbf {\bibinfo {volume} {80}}
  (\bibinfo {year} {2007})}\BibitemShut {NoStop}%
\bibitem [{\citenamefont {Pandey}\ and\ \citenamefont
  {Schumacher}(2020)}]{PandeySchumacher2020}%
  \BibitemOpen
  \bibfield  {author} {\bibinfo {author} {\bibfnamefont {S.}~\bibnamefont
  {Pandey}}\ and\ \bibinfo {author} {\bibfnamefont {J.}~\bibnamefont
  {Schumacher}},\ }\bibfield  {title} {\bibinfo {title} {Reservoir computing
  model of two-dimensional turbulent convection},\ }\href@noop {} {\bibfield
  {journal} {\bibinfo  {journal} {arXiv:2001.10280 [physics]}\ } (\bibinfo
  {year} {2020})}\BibitemShut {NoStop}%
\bibitem [{\citenamefont {Eswaran}\ and\ \citenamefont
  {Pope}(1988)}]{EswaranPope1988}%
  \BibitemOpen
  \bibfield  {author} {\bibinfo {author} {\bibfnamefont {V.}~\bibnamefont
  {Eswaran}}\ and\ \bibinfo {author} {\bibfnamefont {S.~B.}\ \bibnamefont
  {Pope}},\ }\bibfield  {title} {\bibinfo {title} {An examination of forcing in
  direct numerical simulations of turbulence},\ }\href@noop {} {\bibfield
  {journal} {\bibinfo  {journal} {Comp. \& Fluids}\ }\textbf {\bibinfo {volume}
  {16}},\ \bibinfo {pages} {257} (\bibinfo {year} {1988})}\BibitemShut
  {NoStop}%
\bibitem [{\citenamefont {Linkmann}\ and\ \citenamefont
  {Morozov}(2015)}]{LinkmannMorozov2015}%
  \BibitemOpen
  \bibfield  {author} {\bibinfo {author} {\bibfnamefont {M.~F.}\ \bibnamefont
  {Linkmann}}\ and\ \bibinfo {author} {\bibfnamefont {A.}~\bibnamefont
  {Morozov}},\ }\bibfield  {title} {\bibinfo {title} {Sudden relaminarization
  and lifetimes in forced isotropic turbulence},\ }\href@noop {} {\bibfield
  {journal} {\bibinfo  {journal} {Phys. Rev. Lett.}\ } (\bibinfo {year}
  {2015})}\BibitemShut {NoStop}%
\bibitem [{\citenamefont {Mccomb}\ \emph {et~al.}(2015)\citenamefont {Mccomb},
  \citenamefont {Berera}, \citenamefont {Yoffe},\ and\ \citenamefont
  {Linkmann}}]{McCombEtAl2015}%
  \BibitemOpen
  \bibfield  {author} {\bibinfo {author} {\bibfnamefont {W.~D.}\ \bibnamefont
  {Mccomb}}, \bibinfo {author} {\bibfnamefont {A.}~\bibnamefont {Berera}},
  \bibinfo {author} {\bibfnamefont {S.~R.}\ \bibnamefont {Yoffe}},\ and\
  \bibinfo {author} {\bibfnamefont {M.~F.}\ \bibnamefont {Linkmann}},\
  }\bibfield  {title} {\bibinfo {title} {Energy transfer and dissipation in
  forced isotropic turbulence},\ }\href@noop {} {\bibfield  {journal} {\bibinfo
   {journal} {Phys. Rev. E}\ }\textbf {\bibinfo {volume} {91}} (\bibinfo {year}
  {2015})}\BibitemShut {NoStop}%
\bibitem [{\citenamefont {Shih}\ \emph {et~al.}(2016)\citenamefont {Shih},
  \citenamefont {Hsieh},\ and\ \citenamefont {Goldenfeld}}]{ShihEtAl2015}%
  \BibitemOpen
  \bibfield  {author} {\bibinfo {author} {\bibfnamefont {H.~Y.}\ \bibnamefont
  {Shih}}, \bibinfo {author} {\bibfnamefont {T.~L.}\ \bibnamefont {Hsieh}},\
  and\ \bibinfo {author} {\bibfnamefont {N.}~\bibnamefont {Goldenfeld}},\
  }\bibfield  {title} {\bibinfo {title} {Ecological collapse and the emergence
  of travelling waves at the onset of shear turbulence},\ }\href@noop {}
  {\bibfield  {journal} {\bibinfo  {journal} {Nat. Phys.}\ }\textbf {\bibinfo
  {volume} {12}},\ \bibinfo {pages} {245} (\bibinfo {year} {2016})}\BibitemShut
  {NoStop}%
\bibitem [{\citenamefont {Barkley}\ \emph {et~al.}(2015)\citenamefont
  {Barkley}, \citenamefont {Song}, \citenamefont {Mukund}, \citenamefont
  {Lemoult}, \citenamefont {Avila},\ and\ \citenamefont {Hof}}]{BarkleyEtAl}%
  \BibitemOpen
  \bibfield  {author} {\bibinfo {author} {\bibfnamefont {D.}~\bibnamefont
  {Barkley}}, \bibinfo {author} {\bibfnamefont {B.}~\bibnamefont {Song}},
  \bibinfo {author} {\bibfnamefont {V.}~\bibnamefont {Mukund}}, \bibinfo
  {author} {\bibfnamefont {G.}~\bibnamefont {Lemoult}}, \bibinfo {author}
  {\bibfnamefont {M.}~\bibnamefont {Avila}},\ and\ \bibinfo {author}
  {\bibfnamefont {B.}~\bibnamefont {Hof}},\ }\bibfield  {title} {\bibinfo
  {title} {The rise of fully turbulent flow},\ }\href@noop {} {\bibfield
  {journal} {\bibinfo  {journal} {Nature}\ }\textbf {\bibinfo {volume} {526}},\
  \bibinfo {pages} {550} (\bibinfo {year} {2015})}\BibitemShut {NoStop}%
\bibitem [{\citenamefont {Rogallo}(1981)}]{rogallo}%
  \BibitemOpen
  \bibfield  {author} {\bibinfo {author} {\bibfnamefont {R.~S.}\ \bibnamefont
  {Rogallo}},\ }\bibfield  {title} {\bibinfo {title} {Numerical experiments in
  homogeneous turbulence},\ }\href@noop {} {\bibfield  {journal} {\bibinfo
  {journal} {NASA Tech. Memo. 81315}\ } (\bibinfo {year} {1981})}\BibitemShut
  {NoStop}%
\bibitem [{\citenamefont {Donzis}(2007)}]{donzisphd}%
  \BibitemOpen
  \bibfield  {author} {\bibinfo {author} {\bibfnamefont {D.~A.}\ \bibnamefont
  {Donzis}},\ }\emph {\bibinfo {title} {Scaling of turbulence and turbulent
  mixing using Terascale numerical simulations}},\ \href@noop {} {Ph.D. thesis}
  (\bibinfo {year} {2007})\BibitemShut {NoStop}%
\bibitem [{\citenamefont {Yeung}\ \emph {et~al.}(2018)\citenamefont {Yeung},
  \citenamefont {Sreenivasan},\ and\ \citenamefont {Pope}}]{YeungEtAl2018}%
  \BibitemOpen
  \bibfield  {author} {\bibinfo {author} {\bibfnamefont {P.~K.}\ \bibnamefont
  {Yeung}}, \bibinfo {author} {\bibfnamefont {K.~R.}\ \bibnamefont
  {Sreenivasan}},\ and\ \bibinfo {author} {\bibfnamefont {S.~B.}\ \bibnamefont
  {Pope}},\ }\bibfield  {title} {\bibinfo {title} {Effects of finite spatial
  and temporal resolution in direct numerical simulations of incompressible
  isotropic turbulence},\ }\href@noop {} {\bibfield  {journal} {\bibinfo
  {journal} {Phys. Rev. Fluids}\ }\textbf {\bibinfo {volume} {3}} (\bibinfo
  {year} {2018})}\BibitemShut {NoStop}%
\bibitem [{\citenamefont {Buaria}\ \emph {et~al.}(2019)\citenamefont {Buaria},
  \citenamefont {Pumir}, \citenamefont {Bodenschatz},\ and\ \citenamefont
  {Yeung}}]{BuariaEtAl2019}%
  \BibitemOpen
  \bibfield  {author} {\bibinfo {author} {\bibfnamefont {D.}~\bibnamefont
  {Buaria}}, \bibinfo {author} {\bibfnamefont {A.}~\bibnamefont {Pumir}},
  \bibinfo {author} {\bibfnamefont {E.}~\bibnamefont {Bodenschatz}},\ and\
  \bibinfo {author} {\bibfnamefont {P.~K.}\ \bibnamefont {Yeung}},\ }\bibfield
  {title} {\bibinfo {title} {Extreme velocity gradients in turbulent flows},\
  }\href@noop {} {\bibfield  {journal} {\bibinfo  {journal} {New J. Phys.}\
  }\textbf {\bibinfo {volume} {21}},\ \bibinfo {pages} {043004} (\bibinfo
  {year} {2019})}\BibitemShut {NoStop}%
\bibitem [{\citenamefont {Dhruva}\ \emph {et~al.}(1997)\citenamefont {Dhruva},
  \citenamefont {Tsuji},\ and\ \citenamefont {Sreenivasan}}]{Dhruva1997}%
  \BibitemOpen
  \bibfield  {author} {\bibinfo {author} {\bibfnamefont {B.}~\bibnamefont
  {Dhruva}}, \bibinfo {author} {\bibfnamefont {Y.}~\bibnamefont {Tsuji}},\ and\
  \bibinfo {author} {\bibfnamefont {K.~R.}\ \bibnamefont {Sreenivasan}},\
  }\bibfield  {title} {\bibinfo {title} {Transverse structure functions in
  high-{R}eynolds-number turbulence},\ }\href@noop {} {\bibfield  {journal}
  {\bibinfo  {journal} {Phys. Rev. E}\ }\textbf {\bibinfo {volume} {56}},\
  \bibinfo {pages} {R4928} (\bibinfo {year} {1997})}\BibitemShut {NoStop}%
\bibitem [{\citenamefont {Stolovitzky}\ \emph {et~al.}(1993)\citenamefont
  {Stolovitzky}, \citenamefont {Sreenivasan},\ and\ \citenamefont
  {Juneja}}]{StolovitzkyEtAl1993}%
  \BibitemOpen
  \bibfield  {author} {\bibinfo {author} {\bibfnamefont {G.}~\bibnamefont
  {Stolovitzky}}, \bibinfo {author} {\bibfnamefont {K.~R.}\ \bibnamefont
  {Sreenivasan}},\ and\ \bibinfo {author} {\bibfnamefont {A.}~\bibnamefont
  {Juneja}},\ }\bibfield  {title} {\bibinfo {title} {Scaling functions and
  scaling exponents in turbulence},\ }\href@noop {} {\bibfield  {journal}
  {\bibinfo  {journal} {Phys. Rev. E}\ }\textbf {\bibinfo {volume} {48}},\
  \bibinfo {pages} {R3217} (\bibinfo {year} {1993})}\BibitemShut {NoStop}%
\bibitem [{\citenamefont {Watanabe}\ and\ \citenamefont
  {Gotoh}(2007)}]{WatanabeGotoh2007}%
  \BibitemOpen
  \bibfield  {author} {\bibinfo {author} {\bibfnamefont {T.}~\bibnamefont
  {Watanabe}}\ and\ \bibinfo {author} {\bibfnamefont {T.}~\bibnamefont
  {Gotoh}},\ }\bibfield  {title} {\bibinfo {title} {Inertial-range
  intermittency and accuracy of direct numerical simulation for turbulence and
  passive scalar turbulence},\ }\href@noop {} {\bibfield  {journal} {\bibinfo
  {journal} {J. Fluid Mech.}\ }\textbf {\bibinfo {volume} {590}},\ \bibinfo
  {pages} {117} (\bibinfo {year} {2007})}\BibitemShut {NoStop}%
\bibitem [{\citenamefont {Yeung}\ \emph {et~al.}(2015)\citenamefont {Yeung},
  \citenamefont {Zhai},\ and\ \citenamefont {Sreenivasan}}]{YeungEtAl2015}%
  \BibitemOpen
  \bibfield  {author} {\bibinfo {author} {\bibfnamefont {P.~K.}\ \bibnamefont
  {Yeung}}, \bibinfo {author} {\bibfnamefont {X.~M.}\ \bibnamefont {Zhai}},\
  and\ \bibinfo {author} {\bibfnamefont {K.~R.}\ \bibnamefont {Sreenivasan}},\
  }\bibfield  {title} {\bibinfo {title} {Extreme events in computational
  turbulence},\ }\href@noop {} {\bibfield  {journal} {\bibinfo  {journal}
  {Proc. Nat. Acad. Sci.}\ }\textbf {\bibinfo {volume} {112}} (\bibinfo {year}
  {2015})}\BibitemShut {NoStop}%
\bibitem [{\citenamefont {Khurshid}(2021)}]{khurshidthesis}%
  \BibitemOpen
  \bibfield  {author} {\bibinfo {author} {\bibfnamefont {S.}~\bibnamefont
  {Khurshid}},\ }\emph {\bibinfo {title} {Signatures of Fully Developed
  Turbulence and Their Emergence in Direct Numerical Simulations}},\ \href@noop
  {} {Ph.D. thesis} (\bibinfo {year} {2021})\BibitemShut {NoStop}%
\bibitem [{\citenamefont {Friedrich}\ \emph {et~al.}(2018)\citenamefont
  {Friedrich}, \citenamefont {Margazoglou}, \citenamefont {Biferale},\ and\
  \citenamefont {Grauer}}]{FriedrichEtAl2018}%
  \BibitemOpen
  \bibfield  {author} {\bibinfo {author} {\bibfnamefont {J.}~\bibnamefont
  {Friedrich}}, \bibinfo {author} {\bibfnamefont {G.}~\bibnamefont
  {Margazoglou}}, \bibinfo {author} {\bibfnamefont {L.}~\bibnamefont
  {Biferale}},\ and\ \bibinfo {author} {\bibfnamefont {R.}~\bibnamefont
  {Grauer}},\ }\bibfield  {title} {\bibinfo {title} {Multiscale velocity
  correlations in turbulence and {B}urgers turbulence: Fusion rules, {M}arkov
  processes in scale, and multifractal predictions},\ }\href@noop {} {\bibfield
   {journal} {\bibinfo  {journal} {Phys. Rev. E}\ }\textbf {\bibinfo {volume}
  {98}} (\bibinfo {year} {2018})}\BibitemShut {NoStop}%
\end{thebibliography}%
\end{document}